\documentstyle[prl,aps,epsfig,floats,twocolumn]{revtex} 
\def\ni{\noindent}
\def\sig{{\sigma}}
\def\br{\vec{r}}

\def\bs{\vec{s}}
\def\bt{\vec{t}}
\def\bu{\vec{u}}
\def\bn{\vec{n}}

\def\bR{\vec{R}}

\def\bff{\vec{f}}
\def\bFF{\vec{F}}
\def\brho{\vec{\rho}}
\def\bomega{\vec{\omega}}
\def\bnabla{\vec{\nabla}}

\def\hC{{\hat C}}

\def\hP{{\hat P}}
\def\hG{{\hat G}}
\def\hQ{{\hat Q}}

\def\hS{{\hat S}}

\def\hsig{{\hat \sigma}}
\def\heps{{\hat \epsilon}}

\def\cA{{\cal A}}

\def\halfspace{\hskip0.4cm} 
 
\begin{document} 
\draft 
 
 
\title{Stress in planar cellular solids and isostatic granular assemblies: \\ Coarse-graining the constitutive equation}
\author{Raphael Blumenfeld}
 
\address{Polymers and Colloids, Cavendish Laboratory, Madingley Road,  
Cambridge CB3 0HE, UK} 
\maketitle 
\date{\today} 
\maketitle

\begin{abstract} 

A recent theory for stress transmission in isostatic granular and cellular systems predicts a constitutive equation that couples the stress 
field to the local microstructure \cite{BBi}. The theory could not be applied to macroscopic systems because the constitutive equation 
becomes trivial upon straightforward coarse-graining. This problem is resolved here for arbitrary planar structures. The solution is based on 
the observation that staggered order makes it possible to couple the stress to a reduced geometric tensor that can be coarse-grained.  
The method proposed here makes it possible to apply this idea to realistic systems whose staggered order is generally 'frustrated'. This is 
achieved by a renormalization procedure which removes the frustration and enables the use of the upscalable reduced tensor. As an example we calculate the stress due to a defect in a periodic solid foam. 

\end{abstract} 

\narrowtext

Particulate and cellular materials play a major role in everyday life and are relevant to a wide range of technological applications. Yet, the
fundamentals of stress transmission in these systems are not fully understood. Recent work has shown that stress transmission in both cellular 
and isostatic granular systems differs significantly from conventional solids \cite{pw}-\cite{Bi}. All solids in mechanical equilibrium satisfy 
force and torque balance

\begin{equation}
\bnabla\cdot\hsig + F_{ext} = 0 \ \ \ \ \ ; \ \ \ \ \ \hsig = \hsig^T 
\label{eq:Ai}
\end{equation}
where $\hsig$ is the stress tensor, $\hsig^T$ its transpose and $F_{ext}$ is an external loading. To the balance equations one must add a set
of constitutive equations to uniquely determine the stress field. In elasticity theory these equations involve stress-strain relations but this
information is redundant for granular \cite{BBi}-\cite{EG} and cellular systems \cite{Bi}. This is because these systems are isostatic and the  
intergranular and intercellular forces can be in principle determined from statics alone. Note that this is true regardless of material compliance. It has been shown that in 2D the conventional constitutive relation can be replaced by an equation of the form

\begin{equation} 
\sum_{ij} q^{ij}\sig_{ij} = 0 
\label{eq:Aii} 
\end{equation} 
with $\hQ$ a tensor that depends only on the local geometry. Several empirical \cite{MECetal} and mean-field \cite{EG} proposals have been put
forward for the macroscopic form of $\hQ$ and a recent theory gave the specific dependence of this tensor on microstructural 
details \cite{BBi}. In a further development an exact mapping was found between trivalent cellular solids and isostatic granular systems which 
enabled to extend the theory to solid foams \cite{Bi}. For brevity, the following will be discussed in terms of the 
latter. The tensor $\hQ$ is symmetric and is defined as

\begin{figure}
\centerline{\psfig{file=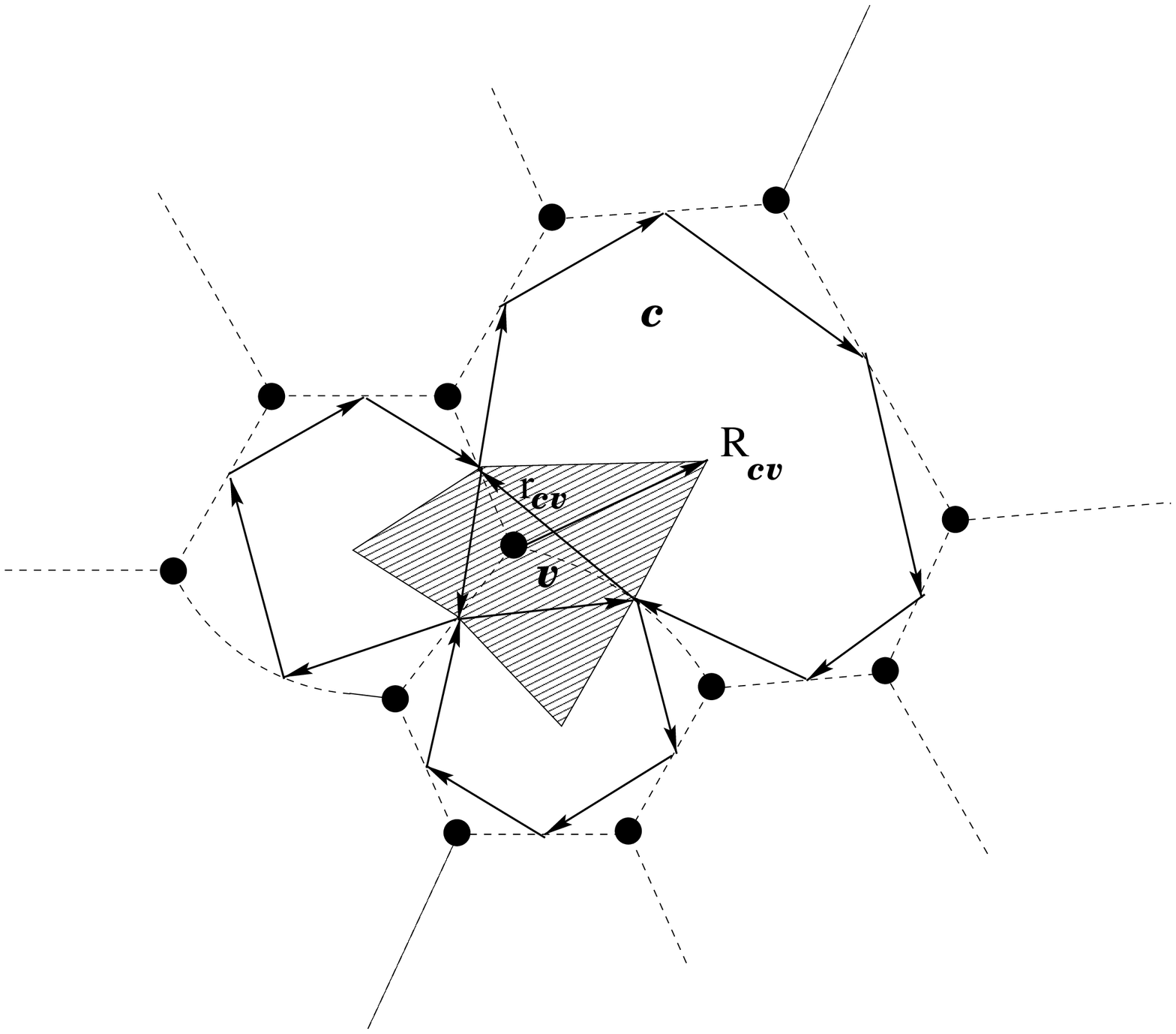,height=6.5cm}}
\caption{The vectors $\br_{cv}$ and $\bR_{cv}$ shared between vertex $v$ and cell $c$. $\br_{cv}$ connects two neighboring wall midpoints and is one edge in a clockwise-directed triangle around vertex $v$. The vector $\bR_{cv}$ points from the center of triangle $v$ to the center of cell $c$.}
\end{figure}
\begin{equation}
\hQ = {1\over 2} \heps^{-1}\left(\hC_v + \hC_v^T\right)\heps
\label{eq:Aiia}
\end{equation}
where $\matrix{\heps}={0 \,\ 1 \choose -1 \,0 }$ is a $\pi/2$-rotation matrix and  

\begin{equation}
C_v^{ij} = \sum_c r^i_{cv} R^j_{cv} \ .
\label{eq:Aiix}
\end{equation}
Here $i, j$ are Cartesian components, $v$ denotes a vertex of the cellular solid, the definitions of the vectors $\br_{cv}$ and $\bR_{cv}$ are
given in fig. 1, and the sum runs over all the cells $c$ that surround vertex $v$. The only contribution to $\hQ$ comes from the parallel
parts of $\bR_{cv}$ and $\br_{cv}$ and therefore $\hQ$ is a measure of the net rotation of the triangle around vertex $v$ relative to its
immediate environment. It is also the net deviation of the quadrilaterals formed by the self-dual $\br$-$\bR$ pairs (see fig. 2) from rhombi
in a given direction. The geometric interpretation of $\hQ$ can be best seen in terms of $\hP = \heps \hQ \heps^{-1}$. Using vector algebra, 
$\hP$ can be rewritten in terms of the vectors $\bs_{cv}$ and $\bt_{cv}$ of fig. 2 as

\begin{equation}
\hP_v = {1\over 2}\sum_c \left( \bs_{cv}\bs_{cv} - \bt_{cv}\bt_{cv}\right) \ .
\label{eq:Aiii}
\end{equation}
The antisymmetric part of each term in $\hC_v$ can be written as $A_{cv}\heps$ where $A_{cv}$ is the area of the quadrilateral whose diagonals 
are $\bR_{cv}$ and $\br_{cv}$. The area associated with vertex $v$ is then $A_v=\sum_c A_{cv}$ and the total area of the system is exactly 
$A_{sys}=\sum_v A_v$.

\begin{figure}
\centerline{\psfig{file=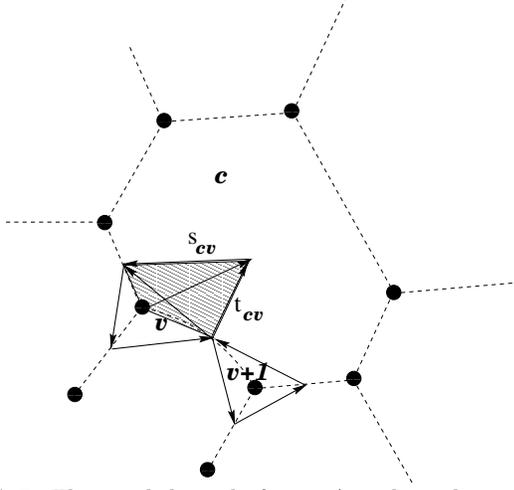,height=6.5cm}}
\caption{The quadrilateral of area $A_{cv}$ whose diagonals are $\br_{cv}$ and $\bR_{cv}$. The vectors $\bs_{cv}$ and $\bt_{cv}$
appear in the symmetric tensor $\hP$. Note that $\bt_{cv}=-\bs_{cv'}$ and therefore a sum over $\hP$ inside cell $c$ vanishes
identically irrespective of the shape of the cell.} 
\end{figure}

Although the first-principles derivation of $\hQ$ encouragingly supports eq. (\ref{eq:Aii}) this equation can only be useful if $\hQ$ can be coarse-grained to yield a macroscopically valid equation. This, however, proved to be non-trivial \cite{BBi} seriously impeding the application of the new theory. This problem is resolved here.

Coarse-graining of constitutive properties has been studied in many contexts. The conventional approach is to integrate over small-scale degrees of freedom, leaving the constitutive properties dependent on large-scale coordinates \cite{coarse-graining}. The procedure is to choose a lengthscale $L$ above which the $d$-dimensional medium appears
statistically homogeneous. The entire system is then regarded as made of basic units of volume $L^d$ and the constitutive property is averaged
within a unit, giving a mean characteristic value. This value is then used in the constitutive equation, which is subsequently regarded as applicable to scales larger than $L$. 
This straightforward approach fails for the constitutive eq. (\ref{eq:Aii}) because the mean value of the tensor $\hQ$ vanishes, giving a trivial 
identity for macroscopic scales. To see this consider a part of the system $\Gamma$ of area $A_{\Gamma}$ and boundary $\partial\Gamma$, which 
consists of $N_{\Gamma}$ vertices. The average of the tensor $\hP$ in $\Gamma$ is

\begin{equation}
\hP_{\Gamma} = {1\over {2A_{\Gamma}}} \sum_{c,v \in \Gamma} \left( \bs_{cv}\bs_{cv} - \bt_{cv}\bt_{cv} \right) 
\label{eq:Aiv}
\end{equation}
From fig. 2 we see that if $v$ and $v'$ are neighbor vertices then $\bt_{cv}=-\bs_{cv'}$ in the cells $c$ that they both border. This means that the terms in sum (\ref{eq:Aiv}) cancel out in pairs for cells that are fully enclosed in $\Gamma$ and only terms along the boundary $\partial\Gamma$ contribute. It follows that $\hP_{\Gamma}$ diminishes with size as $1/\sqrt{N_\Gamma}$. Therefore a simple area average of $\hQ$ vanishes on large scales irrespective of the microstructural characteristics. To use eq. (\ref{eq:Aii}) for macroscopic systems and keep $\hQ$ as the correct geometrical decsriptor we need to develop a coarse-graining method that goes beyond straightforward area averaging. 

The key to the resolution of this problem comes from the observation \cite{BBi} that in systems possessing a staggered order (stagger-ordered systems, SOS) eq. (\ref{eq:Aii}) can be rewritten in terms of only half the degrees of freedom.  A staggered order arises when it
is possible to label all the vertices + and -, such that each vertex has neighbors of only the opposite sign. This requires that all the cells
in the network have an even number of edges. Such networks can be partitioned into +/- vertex pairs, each of area $A_{pair} = A_v^+ + A_v^-$.
The local force moment on the triangle around $v$ is $\hat{S}_v = \sum_{v'}\br_{vv'}\times\bff_{vv'}$, where $v'$ are the neighbors of $v$,
$\bff_{vv'}$ the forces that triangles $v'$ exert on $v$, and $\br_{vv'}$ are the position vectors of the contacts between $v$ and $v'$. By
definition, the local stress at vertex $v$ is $\hsig_v = \hat{S}_v / A_v$. Let us define the mean stress and the stress difference of a pair,
respectively, as $\hsig_m = (S^+ + S^-)/A_{pair}$ and $\hsig_d = (S^+ - S^-)/A_{pair}$. $\hsig_m$ is the stress field, coarse-grained over a scale of two vertices and $\hsig_d$ is the local fluctuation whose average must vanish on larger scales. A manipulation of 
eqs. (\ref{eq:Aii}) for both the + and - vertices gives

\begin{eqnarray}
(\hQ^+ + \hQ^-):\hsig_m + (\hQ^+ - \hQ^-):\hsig_d & = & 0 \nonumber \\
(\hQ^+ - \hQ^-):\hsig_m + (\hQ^+ + \hQ^-):\hsig_d & = & 0 \nonumber
\end{eqnarray}
In an area average of these equations the term linear in $(\hQ^+ + \hQ^-)$ vanishes, as discussed above, and $\hsig_d$ vanishes from its
definition. This renders the first equation a trivial identity. It also means that the second term in the second equation consists of a correlation of 
two small quantities and is therefore negligible relative to the first term. Thus only the first term of the second equation
survives the averaging. Furthermore, the vanishing of the average of $\hQ^+ + \hQ^-$ means that the average of $\hQ^+ - \hQ^-$ is equal to 
that of $2\hQ^+$. Thus, on a lengthscale of several cells

\begin{equation}
\hQ^+ :\hsig_m = 0
\label{eq:Diii}
\end{equation}
This relation has the same form as the original constitutive equation and is therefore a coarse-grained version of it. Its advantage is that, unlike $\hQ$, an area average of $\hQ^+$ {\it does not vanish identically} \cite{vanish} and so the problem of coarse-graining is resolvable in SOS.

Unfortunately, most foams are not SOS and cannot be conveniently partitioned into +/- vertex pairs. The staggered order breaks around cells with
odd numbers of edges (odd-edged cells, OECs) which cannot perfectly accommodate alternating signs around them. At these locations the system is 'frustrated' much like a
system of antiferromagnetically coupled Ising spins.
\begin{figure}
\centerline{\psfig{file=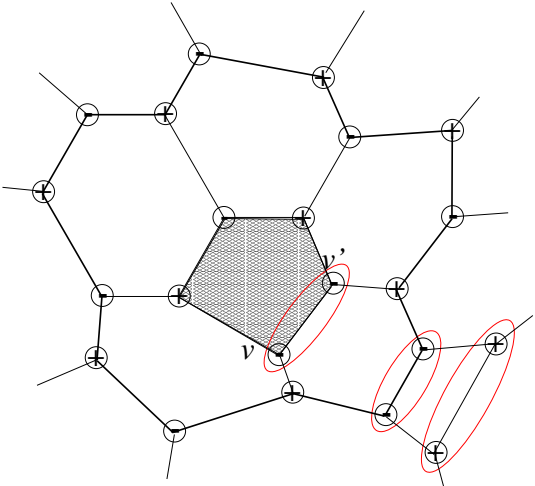,height=6.5cm}}
\caption{The frustration around an OEC (shaded) surrounded by even-edged cells. Labeling the vertices around it alternately by + and -
leaves two neighboring 'frustrated' vertices with the same sign ($v$ and $v'$). Starting from a neighbor of $v$ and labeling the next shell (thick lines), leaves another frustrated pair lying adjacent to the first one. Continuing this procedure results in a line of such frustrated vertex pairs emanating from the OEC, the first three of which are shown. This line can be capped only by encountering another OEC along its path.}
\end{figure}
To resolve the coarse-graining problem for general systems we then need to address the issue of frustration. 

Let us consider first one OEC
embedded in a region of cells with only of even numbers of edgeds (e.g. fig. 3). Starting from an arbitrary vertex, $v$, we label the
vertices alternatingly + and - around the OEC in the clockwise direction. The last vertex, $v'$, has the same sign as the first vertex
$v$. Apply now the same procedure to the first shell of cells surrounding the OEC, starting from the neighbor of $v$ and going clockwise, as shown in fig. 3.
Since there are only even-edged cells around the OEC then this shell must also contain a frustrated pair of vertices and, by construction,
this pair lies adjacent to $v$-$v'$. Repeating this process shell by shell outwards results in a line of frustrated pairs of vertices
emanating from the OEC. Three pairs along this line are shown in fig. 3 highlighted by ellipses. This line of frustrated pairs extends to the boundary of the
system unless it encounters another OEC along its path, in which case it terminates.

This suggests a procedure to isolate the frustrated pairs in general systems and remove their effect: 
1. Identify all the OECs in the network; 
2. partition the OECs into nearest pairs; 
3. following the above procedure, draw a line of frustrated vertex pairs between each pair of OECs avoiding intersection between
lines. Except along these lines, the system is ordered, i.e. it can be completely partitioned into pairs of +/- vertices. The
choice of nearest pairs on the second step minimizes the total number of frustrated pairs. 

\begin{figure}
\centerline{\psfig{file=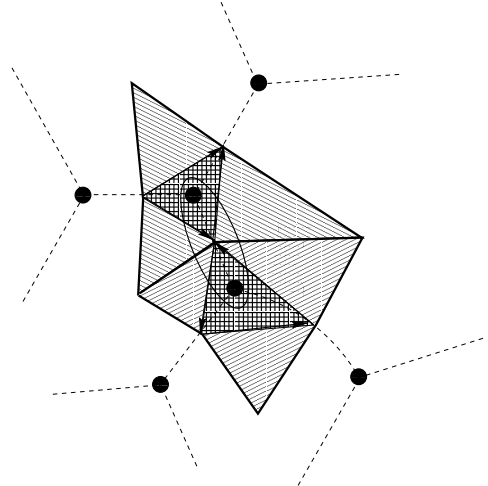,height=6.5cm}}
\caption{Fusion of a frustrated pair of vertices renormalizes away the contact point between the triangles. The area of the renormalized vertex (shaded) is $A_f = A_1 + A_2$, which is the coefficient of the antisymmetric part of the renormalized geometric tensor $\hC_f = \hC_1 + \hC_2$. The common vectors $\bs$ and $\bt$ cancel and therefore the symmteric part of the renormalized grain is independent of the vectors that connect the contact point to the centers of the cells that surround it. Therefore the stress field, that is coupled to the microstructure via the symmetric part of $\hC$ is unaffected by the fusion.}
\end{figure}

The frustration is removed by renormalization. Each frustrated pair of vertices along the frustration lines (say $v$ and $v'$ in fig. 3) is regarded as one supervertex. These vertices (the equivalent of grains) are pressed against each other anyway and therefore this does not alter the original external forces on the pair. The renormalized pair geometric tensor is defined as $\hC_{pair} = \hC_v + \hC_{v'}$ and hence $\hP_{pair} = \hP_v + \hP_{v'}$. Since this tensor determines the stress through eq. (\ref{eq:Aii}) it remains to show that this operation leaves the stress field outside the supervertex intact. Using eq. (\ref{eq:Aiii}) and inspecting fig. 2 note that the vectors $\bs_{cv'}$ and $\bt_{cv}$, which connect the
$v$-$v'$ contact point to the center of cell $c$, cancel out in this sum. So do their counterparts in the opposite cell and so $\hP_{pair}$ is independent of this contact. This also means that the stress tensor becomes independent of $f_{vv'}$. The remaining $\bs$ and $\bt$ vectors surround the supervertex as if it were a regular vertex and the geometric tensor outside the supervertex remains unaffected. Therefore the renormalization operation ineed leaves the stress field outside this pair unchanged, as
required. Another convenient feature of the procedure is that the area of the supervertex is $A_{pair} = A_v + A_{v'}$, exactly corresponding to the sum of the adjacent areas of the two original grains (see fig. 4) and so the volume of the system is preserved. In granular packings this procedure is the same as actual fusion of the pair of grains. Once all the frustrated pairs are renormalized the system becomes fully SOS and the above coarse-graining can be used. This procedure then makes it possible to upscale the constitutive eq. (\ref{eq:Aii}) in general systems.

To illustrate the method, consider the structure shown in fig. 5. It is a honeycomb lattice in which one cell has been made into an 
octagon at the expense of two neighbors that are reduced to pentagons. Except around the pentagons the structure is SOS with $\hP$ vanishing due to the hexagonal symmetry. The structural defect gives rise to a fluctuation in the constitutive properties, which we wish to determine. Following the procedure outlined above, we renormalize the two pentagonal cells by fusing the two pairs enclosed by ellipses. Summing over the $\hQ$-tensors of the positive vertices gives
$\matrix{\hQ_{defect}}={-\frac{3}{10} \,\ -\frac{4\sqrt{3}}{5} \choose -\frac{4\sqrt{3}}{5} \,-\frac{3}{2} }$ and, after some algebra, the constitutive equation reads

\begin{equation}
\sig_{11} + 5\sig_{22} + \frac{16}{\sqrt{3}}\sig_{12} = 0
\label{eq:Div}
\end{equation}
Combined with eqs. (\ref{eq:Ai}), this relation gives the stress on the scale of the entire defect.
\begin{figure}
\centerline{\psfig{file=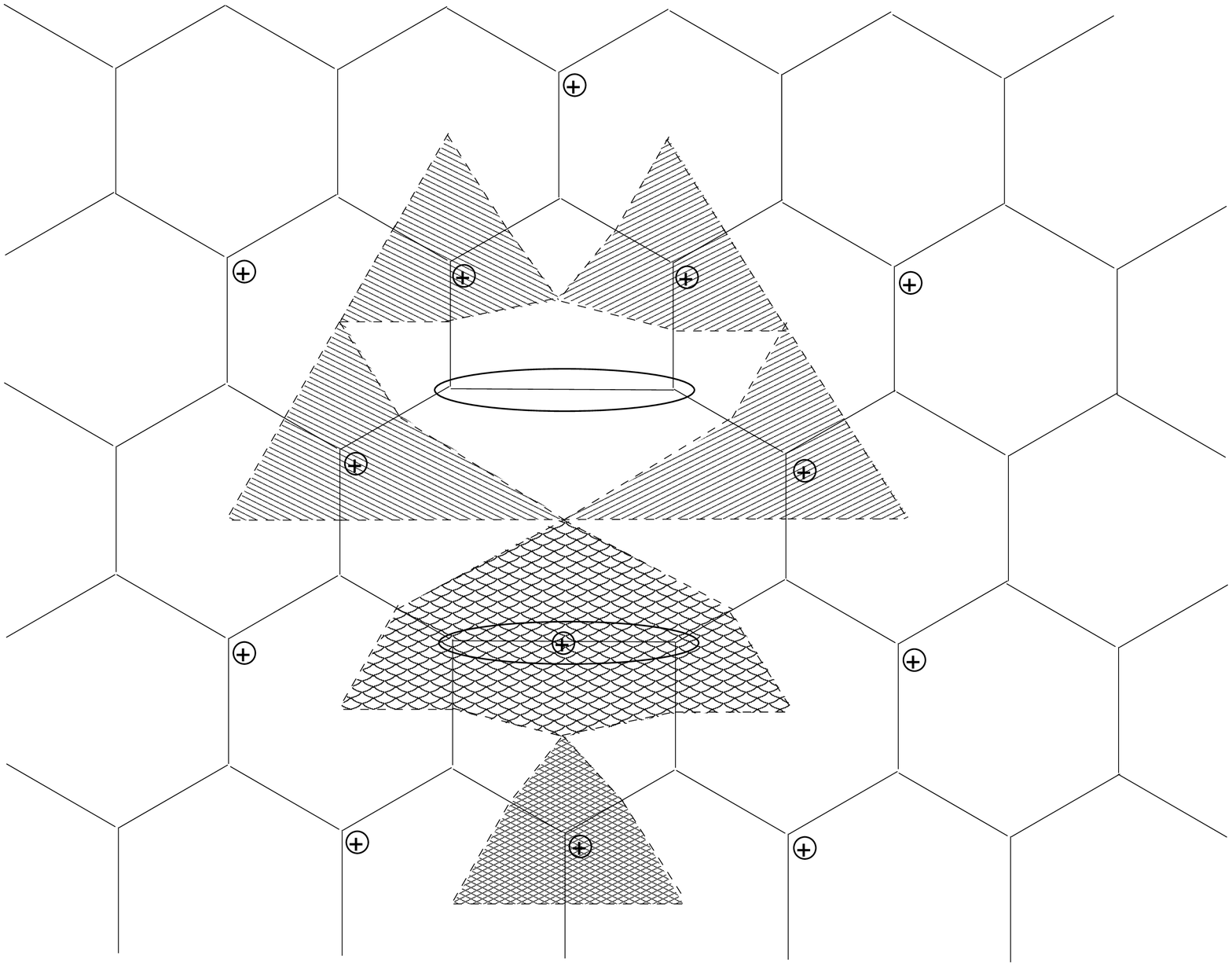,height=6.5cm}}
\caption{A honeycomb cellular structure containing a defect which consists of an octagonal cell straddled by two pentagonal 
cells. The coarse graining is over all the positive vertices but only those near the defect with irregular areas (shown 
shaded) have a finite $\hP$.}
\end{figure}

To conclude, this communication has addressed the application of the microscopic theory for stress transmission in trivalent solid foams and marginally rigid granular assemblies to macroscopic lengthscales. Under conventional coarse-graining the equation that couples the stress field to the microstructure becomes uselss an upscaling method has been proposed to overcome this problem. The method consists of fusion of frustrated vertices / grains, leading to a renormalized system that is fully stagger-ordered and whose structure is characterized by a suitably renormalized geometric tensor. The renormalized system can be partitioned into two subsystems, much like antiferromagnetic sublattices, and this enables coarse-graining of the constitutive equation. The method makes it possible to extend the microscopic theory to macroscopic systems, paving the way for calculations that are useful for engineering purposes. An example of a localized defect in an otherwise perfect honeycomb lattice has been solved explicitly. This solution can be used to give the Green function for the stress field in generally disordered lattices but this issue will be discussed in more detail elsewhere. Since the microscopic theory does not resort to stress-strain relations then neither does the upscaled macroscopic version. This obviates the problem of the structure-property relationship as we know it in these systems. Specifically, there is no need to take the conventional approach where a relationship is sought between the local structure and elastic constants, which are subsequently coarse-grained. Rather, the problem reduces to the direct coarse-graining of the structural tensorial descriptor $\hQ$.  The new method also makes it straightforward to predict stress concentration in cellular materials and relate the occurrence of 'hot spots' to local microstructural defects. This is expected to improve the prediction of failure in advanced materials as well as to set new goals for material processing methods to manipulate microstructural characteristics in order to avoid such vulnerability.

\vspace{0.5cm}
\ni {\bf Acknowledgements} 
 
\ni I thank Sir S. F. Edwards, R. C. Ball, and M. Schwartz for useful conversations. I am indebted to J. Brujic for critical reading.


\begin{thebibliography}{99}
\bibitem{vanish} Note, however, that this average should vanish on scales where the system is statistically isotropic.
\end{thebibliography}
\end{document}